\documentclass[11pt]{article}  
\usepackage{graphicx}
\usepackage{amssymb,amsmath}
\newcommand{\be}{\begin{equation}}
\newcommand{\ee}{\end{equation}}
\newcommand{\bea}{\begin{eqnarray}}
\newcommand{\eea}{\end{eqnarray}}
\begin{document}
\begin{center} 
{\LARGE{\bf  Particle localization on helical nanoribbons:
Quantum analog of the Coriolis effect}}
\end{center} 
\vspace{.3in}
\begin{center} 
{{\bf Radha Balakrishnan$^{(1)}$, Rossen Dandoloff$^{(2)}$, Victor Atanasov$^{(2)}$ and Avadh Saxena$^{(3)}$}}\\ 
{$^{(1)}$The Institute of Mathematical Sciences, Chennai 600 113, India \\
$^{(2)}$Department of Condensed Matter Physics and Microelectronics, Faculty of Physics, 
Sofia University, 5 Blvd. J. Bourchier, 1164 Sofia, Bulgaria \\ 
$^{(3)}$Theoretical Division and Center for Nonlinear Studies, Los
Alamos National Laboratory, Los Alamos, New Mexico 87545, USA}
\end{center} 

\vspace{.6in}
{\bf {Abstract:}}
We derive  the Schr\"odinger equation for a  particle confined to the surface of a normal  and a binormal helical nanoribbon, obtain the 
 quantum potentials  induced by their respective  curved surface geometries, and study the localized states of the  particle for each ribbon. When
  the particle momentum satisfies a certain 
 geometric condition,  the particle  localizes near the inner edge for a normal ribbon, and on  the  central helix for a binormal ribbon. This  result suggests the presence of  a  pseudo-force  that pushes the particle transversely along the width of the ribbon. We show that this  phenomenon can be interpreted as a quantum analog of the Coriolis effect, which causes a  transverse deflection of a  classical particle moving in a rotating  frame. 
 We invoke Ehrenfest's theorem applicable to localized states and identify the quantized angular velocities of the rotating frames for the two ribbons. If the particle is an electron, its localization at a specific width gives rise to a Hall-like voltage difference across the ribbon's width. However, unlike in the Hall effect, its origin is not an applied magnetic field,  but  the ribbon's  curved  surface geometry. When a normal  helical ribbon  is mechanically flipped to a binormal configuration in a periodic fashion, it results in  a periodic electron transport from the inner edge to the center, giving rise to a quantum AC voltage. This can be used for designing nanoscale electromechanical devices. Quantum transport on a helical nanoribbon  can be controlled by  tuning the  bends and twists of its surface, suggesting diverse applications in  biopolymers and nanotechnology.
 




\newpage 
  
\section{Introduction}
\label{intro}
Helical nanostructures appear 
in  various contexts in nature.  DNA \cite{watson}, alpha-helices as well as 
beta-sheets in proteins \cite{pauling} 
are some well known examples  in biology. 
In addition, with the advent of nanotechnology, the fabrication  of helical structures with 
specific geometries has been attracting the attention of materials scientists in order to create devices such as miniature electromechanical sensors \cite{sensors}. Helically shaped chiral organic molecules called helicenes \cite{tounsi} have been used in designing chemical sensors and  in optical applications.

 In this paper, we focus  on helical ribbons. Such ribbons have been fabricated using 
a variety of  materials  such as
 silicon carbide, silicon oxide, zinc oxide,   carbon (graphene nanoribbons), molybdinum disulphide, tungsten disulphide, etc. \cite{ gao, liu, prevost, fan}. In general, helical ribbons can be broadly classified into  two types,  called {\em normal} and {\em binormal}  helical ribbons  \cite{goriely,fonseca, zhan}. These differ in their surface geometries. Interestingly, using Kirchhoff's  model for helical strips, it has been shown \cite{goriely}  that  only these two types of helical strips represent consistent and dynamically stable configurations.
 
 To  understand the  transport properties of a particle on
  a helical ribbon of {\em nanodimensions}, it is imperative to study its {\em quantum} behavior on the  curved  surface  of the ribbon. 
 A formulation for studying the quantum mechanics of a particle confined to any {\em general}  curved surface has been given by da Costa \cite{daCosta}. It leads to a  geometric potential  of the form $V=\frac{\hbar^2}{2m}(M^2-K)$ in the Schr\"odinger equation for a particle of mass $m$. Here, $M$ and $K$ represent, respectively, the mean and Gaussian curvatures \cite{struik} of the surface concerned. These arise  essentially due to the embedding of the  2D curved surface in 3D space. In this paper, we apply this  formulation to  the surfaces of normal and binormal  helical nanoribbons,   with a view to predict their respective surface curvature  effects on quantum particle transport.

  Our methodology is as follows: As is well known, a curve in 3D space can be represented in terms of its curvature and torsion, by using the Frenet-Serret equations \cite{struik}. These describe the rotation of  the frame made up of the orthogonal unit triad of vectors  consisting of its local tangent ${\bf t}$, the normal  ${\bf n}$ and  the binormal ${\bf b}$ vectors,  as one moves along  the curve.  We  begin by constructing  the surfaces of  normal and binormal  helical ribbons, using a central circular helix as the base curve, with their widths along  ${\bf n}$ and ${\bf b}$, respectively.  We then determine 
 the  mean curvature and  the Gaussian curvature \cite{struik}  for the curved surfaces of the  two ribbons. We 
 write down the Schr\"odinger equation for a  particle on these curved surfaces by using the Laplace-Beltrami operator, and obtain 
 the  curved-geometry  induced  potential experienced by the quantum  particle using da Costa's formulation \cite {daCosta}. This potential, which is  purely quantum mechanical in origin, consists of a contribution from  the intrinsic (metric) of the surface, as well as a contribution arising from the  extrinsic term that depends on the mean and Gaussian curvatures. We then analyze the expressions obtained for the effective quantum geometric  potential for the  two helical ribbons. It is seen to  depend on both the curvature and the torsion (i.e., twist) of the basic helix used to construct the ribbon surfaces.

We obtain several interesting results. When the quantum particle momentum along the longitudinal helical direction on the ribbon satisfies a certain geometric condition,  the induced quantum geometric  potential supports  localized states in the transverse direction, i.e., on the width of of the ribbon. We show that  the particle  gets  localized  near the inner edge  for  the  normal  helical ribbon, and  on  the  central helix for the binormal helical ribbon. This result suggests  the presence of  a pseudo-force   which  deflects  the particle transversely 
 along the width of the ribbon, to  cause its localization  at  distinct  width values on the two ribbons.  
 
 If the  particles are electrons, the above  localization  effect leads to
the accumulation of  a negative charge near the inner edge of the normal helical ribbon, and on the central helix for the binormal ribbon. This gives  rise to
a  corresponding  transverse quantum voltage on the ribbon. 
 However, this effect is unlike the well known  Hall effect which  arises only  in the presence of a magnetic field \cite{kittel,yoshioka}.  For the two types of helical ribbons,  it is their distinct curved  surface geometries  that give rise to  the two types of  Hall-like voltages. Similarity  with  the Hall effect suggests the presence of a fictitious `magnetic field' on the ribbon whose origin is in the curved geometry of the ribbon.  Note that the Hall effect without a magnetic field in condensed matter is known as the anomalous Hall effect (AHE) \cite{nagaosa} and its quantum analog as the quantum anomalous Hall effect (QAHE) \cite{chang}. 
  
 Interestingly, it has been noted recently that by twisting layers of tungsten disulfide (WS$_2$) to form a 3D spiral, one  generates an electron-path deflecting effect. It has been termed `twistronic Hall effect' \cite{ji}, since it is a Hall-like effect  observed in the absence of a magnetic field. Moreover, this effect is seen to mirror \cite{agarwal}  the Coriolis effect \cite{goldstein}  that is associated with the deflection of  a moving particle observed  in rotating (noninertial) systems. We parenthetically remark that 
quantum Coriolis force
has  been previously mentioned  in the context of non-Hermitian Hamiltonians \cite{gardas}.
 
 This motivates us to ask whether the surface geometry induced pseudo-force we had alluded to earlier,
  which causes the quantum particle (including an electron) to  localize along the ribbon width, can be interpreted as a quantum analog of the Coriolis force which causes a  transverse deflection of a  classical particle moving in a rotating  frame.
  We answer this question in the affirmative. By invoking Ehrenfest's theorem \cite{messiah} applicable to localized states, we present a detailed analysis to find the expressions for the quantized angular velocities of the rotating frames for the two ribbons, and corroborate this interpretation.
 
 We  discuss  the  case when the length $L$ of the central helix on the helical ribbon is composed of $q$ full $2\pi$ turns. We show that the quantized values of the momentum $\rm{k}$ are given in terms of two integers $n$ and $q$, i.e.,  $\rm{k}={\rm k}_{n,q} =  (n/2q) \sqrt{(k_{0}^2 +\tau_0^2)}$, $n= 1, 2, 3...$. Here $k_{0}$ and 
  $\tau_{0}$ are  the curvature and torsion of the helix, to be defined in the next section. We find that the condition for the existence of  localized states is given by 
$1 \le n^{2} \le  q^{2}$. This gives the interesting result that  for such states, the quantum number $n$ gets restricted by the number  $q$ representing  full turns of the helix.

 Various applications of our results on localized states can be envisaged. For example,  mechanically flipping   a  binormal  helical ribbon  to a normal helical configuration in a periodic fashion  leads to periodic electron transport  between the central helix and the inner edge,  resulting in an AC  voltage. This may be useful in designing electromechanical  sensors \cite{sensors}. If  the nanoribbons are elastic,  then depending on the surface elastic energy and the electronic energy,  injection of electrons on a binormal helical ribbon can twist it  into a normal helical ribbon configuration or vice versa. This phenomenon, which could  possibly arise spontaneously in some environments, may be important in biopolymers. Our results demonstrate explicitly that the  transport of a quantum particle on a helical nanoribbon can be controlled by  tuning the bends and twists of the ribbon surface, suggesting potential applications in nanotechnology. 
\section {Construction of  normal and binormal helical ribbons}
\label{constr}
A circular helix  is represented in terms of its arc length $s$ as
${\bf R}(s) = [ R_0 \cos\alpha s, R_0 \sin \alpha s, (P_0/2 \pi) \alpha s] \,$, where $R_{0}$ is the radius of the helix,  $P_{0}$ is its pitch and  $\alpha = 2 \pi/\sqrt{4 \pi^2 R_0^2 +P_0^2}$.
Thus, ${\bf t} (s)=d{\bf R}/ds = \alpha [-R_0 \sin \alpha s, R_0 \cos \alpha s, P_{0}/2\pi]$ is the {\em unit tangent vector} on the helix. The {\em unit  normal vector }  is defined as
${\bf n}(s) ={\bf t}_{s}/|{\bf t}_{s}|$, and the {\em unit binormal  vector} as ${\bf b}(s) = [{\bf t}(s) \times {\bf n}(s)]$. These 
 lie on a plane perpendicular to the tangent ${\bf t}(s)$.
The  orthonormal unit vector triad   $[{\bf t}, {\bf n}, {\bf b}]$ for a {\em general} curve  satisfies the well known 
Frenet-Serret equations \cite{struik}
\be
{\bf t}_s = \kappa  {\bf n} \,\,; \,\, {\bf n}_s = -\kappa {\bf t}  + \tau {\bf b}\,\,;\,\, {\bf b}_{s} = -\tau {\bf n},
\label{FrS}
\ee
where the subscript $s$ stands for derivative with respect to $s$. In addition, the curvature of the curve is
$\kappa = |{\bf t}_{s}|$, and its torsion $\tau = [{\bf t}.({\bf t}_{s }\times {\bf t}_{ss})]/ \kappa ^{2}$ which is a measure of the nonplanarity of the curve.  Further, Eqs. (\ref{FrS})  can be combined to give
 $ {\boldsymbol \sigma }_s = {\boldsymbol \omega_{D}}\times  {\boldsymbol\sigma }$,
 where  ${\boldsymbol \sigma}$ stands for ${\bf t}, {\bf n}$ or ${\bf b}$, and ${\boldsymbol \omega}_{D} =\tau {\bf t}+k {\bf b}$ is called the Darboux vector. Thus the Frenet frame   made up of the orthogonal unit triad  
 [${\bf t}, {\bf n}, {\bf b}$]  {\em rotates} with an angular velocity ${\boldsymbol \omega}_{D}$, as one moves along the curve parametrized by $s$. 

Using the definitions given below Eq. (\ref{FrS}),  we  find the curvature  $\kappa$ and the torsion $\tau$ for the helix to be
$\kappa= k_0 = \alpha^2 R_0$ and $\tau = \tau_{0} = \alpha^2 P_0/2\pi $. These yield $k_{0}/\tau_{0}= 2\pi R_{0}/P_{0}$, and 
\be
\alpha = \sqrt{(k_{0}^2 +\tau_0^2)}.
\label{alpha}
\ee
Hence, for a helix  the curvature and torsion are constants determined by its radius and its pitch.

It is convenient to write all the relevant quantities for the helix in terms of $k_{0}$ and $\tau_{0}$  using the above relationships. Thus the  parametric equation 
  for the helix  becomes
 \be 
 {\bf R}(s) =  (1/\alpha^{2})[ k_0 \cos\alpha s, k_0 \sin \alpha s, \tau_{0}\alpha s].
 \label{helix_s}
 \ee 
From   Eq. (\ref{helix_s})  we find
\be
{\bf t} (s)= d{\bf R}(s)/ds =  [-k_{0} \sin \alpha s, k_0 \cos \alpha s, \tau_{0}]/\alpha . 
\label{Vectktau}
\ee
Using Eq. (\ref{Vectktau}) in Eqs. (\ref{FrS}) with $k=k_{0}$ and $\tau=\tau_{0}$, we  find the normal  vector ${\bf n}$ and the binormal vector  
${\bf b}$ for the helix  to be
\be
{\bf n}(s) = {\bf t}_{s}/|{\bf t}_{s}| = [-\cos \alpha s, -\sin \alpha s, 0]\,\,\,\,;\,\,\,\,{\bf b}(s) =
[\tau_{0} \sin \alpha s, -\tau_{0} \cos\alpha s, k_{0}]/\alpha \, , 
\label{Vecnbktau}
\ee
where $\alpha$ is defined in Eq. (\ref{alpha}).  The length of a helix with  integer $q$ turns of $2\pi$ is given by
 \be
 L (q) =  2 \pi q/ \sqrt{(k_{0}^2 +\tau_0^2)},
 \label{Lq}
 \ee
 where $q=1,2,3...$. 

The surfaces of the {\em normal and  binormal  helical ribbons }  are constructed, respectively,  using  the following  position vectors 
${\bf X}^{(n)} (s, \xi)$ and  ${\bf X}^{(b)} (s, \xi)$ parametrized by $s$ and $\xi$:
\be
{\bf X}^{(n)} (s, \xi) = {\bf R}(s) -\xi {\bf n}(s) 
\label{n}
\ee
and 
\be
{\bf X}^{(b)} (s, \xi) = {\bf R}(s) + \xi {\bf b}(s).
\label{b}
\ee
In the above,  ${\bf R}(s)$ is the central helix on the ribbon, given in Eq. (\ref{helix_s}).  The vectors ${\bf n}$ and ${\bf b}$ for this helix were found in
Eq. (\ref {Vecnbktau}).  The arc length
$s = [0,L]$ ($L$ =length of the helix) and  $\xi$ is the parameter along the width of the ribbon (which is in  the local normal and binormal direction, respectively, for the two types of ribbon), with $\xi = [-d,d]$. Here $2d$ is the width of the ribbon. 



\section {Geometric parameters for normal  and binormal helical ribbons}

As is well known \cite{struik}, the geometry of   a 2D surface created by a position vector ${\bf X} (s, \xi)$ 
depends on two quadratic differential forms.
The First fundamental form denoted by I  is just the local metric on the surface given by $I = (d{\bf X})^2 =  E ds^{2} + 2F ds d\xi + G d\xi^{2}$,
where
\be
E = {\bf X}_s\cdot{\bf X}_s\,\,\,;\,\,\, F= {\bf X}_s\cdot{\bf X}_{\xi}\,\,\,;\,\,\, G = {\bf X}_{\xi}\cdot{\bf X}_{\xi}.
\label{EFG}
\ee
 The unit normal ${\bf N}(s,\xi)$ at every point $(s,\xi)$ on  a  surface  ${\bf X}(s, \xi)$ 
is defined as
\be
{\bf N}(s, \xi) = \frac {({\bf X}_{s} \times {\bf X}_{\xi})}{ |({\bf X}_{s} \times {\bf X}_{\xi})|} =   \frac {({\bf X}_{s} \times {\bf X}_{\xi})}{ \sqrt{EG-F^2}} .
\label{N}
\ee
The second fundamental form denoted by II encodes how the 2D surface is embedded in the ambient 3D space.
It is a measure of how the normal vector  to the surface  ${\bf N}$ varies as one moves on the surface. It is given by  $II = -d{\bf X}. d{\bf N} = e ds^2 + 2f ds d\xi + g d\xi^{2}$,
where
\be
e= - {\bf X}_s\cdot {\bf N}_s={\bf X}_{ss}\cdot {\bf N}\,\,\,\,;\,\,\,\, f = - {\bf X}_s\cdot {\bf N}_{\xi} = - {\bf X}_{\xi}\cdot {\bf N}_{s}\,\,\,\,;\,\,\,\, g = -{\bf X}_{\xi}\cdot {\bf N}_{\xi} \,. 
\label{efg}
\ee
Gaussian curvature $K$ and  mean curvature $M$ of the surface are given in terms of $E,F,G,e,f,g$
as \cite{struik}
\be
K =(eg-f^2)/(EG -F^2)\,;\,\, M= (gE-2fF+eG)/ 2(EG - F^2) \,. 
\label{KM}
\ee
Next, we use all the above expressions to obtain the various geometrical parameters for the normal 
and binormal helical ribbons given in Eqs. (\ref{n}) and (\ref{b}). We use the superscripts $(n)$  and $(b)$ to denote quantities pertaining to a {\em normal}  and a {\em binormal} helical ribbon, respectively.

  Differentiating Eqs. (\ref{n}) and (\ref{b}) with respect to $s$ and $\xi$ respectively,  and using the Frenet-Serret equations  (\ref{FrS}) for the helix (with $\kappa=\kappa_0$ and $\tau =\tau_{0}$), we get
\be 
{\bf X}_{s}^{(n)}  = (1 +k_0\xi){\bf t} -\xi \tau_{0} {\bf b}\,;\,\,\,{\bf X}_{\xi}^{(n)}= -{\bf n}\,;\,\,\,{\bf X}_{s}^{(b)}  = {\bf t} -\xi \tau_{0}{\bf n}\,;\,\,\,{\bf X}_{\xi}^{(b)}= {\bf b}.
\label{XnXbDer}
\ee
Using the above in Eq. (\ref {EFG}) , we obtain 
\be
 E^{(n)} (\xi) =[(1+k_0\xi)^2 + \tau_0^{2} \xi^{2}]= [1+ 2 k_0\xi + (k_0^2+\tau_0^2) \xi^2] ;\,\,\,F^{(n)}=0\,\,;\,\,G^{(n)} = 1 
\label{EFGn}
\ee
and
\be
E^{(b)} (\xi) = 1 + \tau_0^{2} \xi^{2}\,\,;\,\,F^{(b)}=0\,\,\,\,\, ;\,\,\,\,\, G^{(b)} = 1.
\label{EFGb}
\ee
Hence the metrics for the normal  and binormal helical ribbons are given by 
$[d{\bf X}^{(n)}]^2 = E^{(n)}(\xi) ds^2  + d \xi^{2}$ and $[d{\bf X}^{(b)}]^2 = E^{(b)}(\xi) ds^2  + d \xi^{2}$, respectively.\\

On using Eq. (\ref{XnXbDer})   in Eq. (\ref{N}), we find the surface normals of the normal and binormal helical ribbons to be
\be 
{\bf N}^{(n)} = [-(1+k_0 \xi){\bf b} -\xi \tau_{0} {\bf t} ]\,/ \sqrt{E^{(n)}(\xi)}\,\,;\,\,{\bf N}^{(b)} = [-{\bf n} -\xi \tau_{0} {\bf t}]\,/ \sqrt{E^{(b)}(\xi)} \,. 
\label{NnNb}
\ee
A short calculation yields
\be
{\bf N}^{(n)}_s=\tau_{0} {\bf n}/\sqrt{E^{(n)}(\xi)};\,\,{\bf N}^{(n)}_{\xi} = [\tau_0^2 \xi  {\bf b} - \tau_{0} (1+k_0 \xi) {\bf t}]/(E^{(n)})^{3/2};\,\,{\bf N}^{(b)}_s = [k_0{\bf t} - \tau_{0} {\bf b}-\xi \tau_{0}k_{0} {\bf n}]\,/ \sqrt{E^{(b)}(\xi)},
\label{NnNbDer}
\ee
while
 ${\bf N}^{(b)}_{\xi}$ is  found to be a vector with components  only along ${\bf t}$ and ${\bf n}$.
Next, using   Eqs. (\ref{XnXbDer}) and (\ref{NnNbDer}) in  Eq. (\ref{efg}) we obtain
\be
e^{(n)}= 0\,\,\,;\,\,\, f^{(n)}  =\tau_{0}/ \sqrt{E^{(n)}} \,\,\,;\,\,\, g^{(n)} =0 
\label{efgn}
\ee
and
\be
e^{(b)}= -k_0 \sqrt{E^{b}(\xi)};\,\,\,f^{(b)} =\tau_{0}/ \sqrt{E^{b}(\xi)}
\,\,\,\,;\,\,\,\, g^{(b)}  =0.
\label{efgb}
\ee

By substituting  Eqs. (\ref{EFGn}) and (\ref{efgn}) for the normal ribbon, and  Eqs. (\ref{EFGb}) and (\ref{efgb})  for the binormal ribbon  respectively in Eq.(\ref{KM}), we obtain the respective expressions  for the Gaussian curvature $K$ and the mean curvature $M$. Using them, the quantity $(M^{2} -K)$ is found. We get
 \be
K^{(n)} = -\tau_0^{2}/ [E^{(n)} (\xi)]^{2} ;\,\,\,M^{(n)} =0;\,\,\,[M^{(n)}]^2 - K^{(n)}= \frac{\tau_0^{2}}{ [E^{(n)} (\xi)]^{2}} 
\label{KnMn}
\ee
 for the normal helical ribbon, and
 \be
K^{(b)} = -\tau_0^{2}/  [E^{(b)} (\xi)]^{2} ;\,M^{(b)} = -k_0/2 \sqrt{E^{(b)}(\xi)};\,\,\,\,[M^{(b)}]^2 - K^{(b)}= \frac{k_{0}^{2}}{4 E^{(b)}(\xi) } +\frac{\tau_{0}^{2}}{(E^{(b)}(\xi) )^{2}}
\label{KbMb}
\ee
for the binormal helical ribbon.

 Since the mean curvature  $M^{(n)}$  vanishes for the  normal helical ribbon,  it represents a minimal surface. The quantity $(M^{2} - K)$
 will play an important role in the study of  quantum  particle  transport on both the helical ribbons.


 \section { Quantum  particle  transport on normal and binormal helical ribbons}
\subsection {Schr\"odinger equation for a  particle on a  general curved 2D surface}
In a seminal paper, da Costa \cite{daCosta} has presented a consistent  formulation for studying the quantum mechanics of a particle constrained to move on a  two dimensional curved surface ${\bf X}(s,\xi)$  embedded in a three dimensional space. A lengthy and detailed analysis yields the following time-independent Schr\"odinger equation for the surface wave function $\chi(s,\xi)$, valid for {\em all surfaces}  with orthogonal 
 curvilinear coordinates, i.e., $F={\bf X}_s.{\bf X}_\xi =0$ in the  surface metric: 
 \be
 -\dfrac{\hbar^2}{2m}\dfrac{1}{h_{1} h_{2}} \Big [\dfrac{\partial}{\partial s}\dfrac{h_{2}}{h_{1} }\dfrac{\partial \chi}{\partial s}  + \dfrac{\partial}{\partial \xi}\dfrac{h_{1}}{h_{2} }\dfrac{\partial \chi}{\partial \xi}\Big]-\dfrac{\hbar^2}{2m}\big[M^2- K\big]\chi = \mathcal{E}\,\,\chi,
 \label{Lame}
 \ee
 where  $h_{1}$, $h_{2}$ are the Lam\'{e}  coefficients. Here $M$ and $K$ are the mean curvature and Gaussian curvature for the surface.
 
 \subsection{Schr\"odinger equation on the surface of  helical ribbons}
 \label{subsecSE}
 As seen from the last entry in  Eqs. (\ref{KnMn})
and (\ref{KbMb}) respectively,  the quantity  $(M^2 - K)$ appearing in Eq. (\ref{Lame}) is expressed in terms of the respective surface metric parameters $E(\xi)$, for {\em both}  normal and binormal ribbons.  In the case of these helical ribbons, it  is therefore convenient to
express the Lam\'{e} coefficients $h_1$ and $h_2$ appearing in Eq. (\ref{Lame}) in terms of $E(\xi)$.
As shown in  Eqs. (\ref{EFGn}) and (\ref{EFGb}),  $F=0$ and $G=1$ for both ribbons. 
Thus  the  surface metric is given by $(d{\bf X})^2= h_{1}^2 ds^{2 } + h_{2}^{2} d\xi^{2}=  E(\xi) ds^{2} + d\xi^{2}$, yielding  $h_{1}= \sqrt {E(\xi)}$ and $h_{2}= 1$ for both ribbons.  
Using these in  Eq. (\ref{Lame}), we obtain
 \be
 -\dfrac{\hbar^2}{2m}\dfrac{1}{\sqrt{E}} \Big [\dfrac{\partial}{\partial s}\dfrac{1}{\sqrt{E} }\dfrac{\partial \chi}{\partial s}  + \dfrac{\partial}{\partial \xi} \sqrt{E}\dfrac{\partial \chi}{\partial \xi}\Big]-\dfrac{\hbar^2}{2m}\big[M^2- K\big]\chi = \mathcal{E} \,\chi .
 \label{se_E}
 \ee
  Note that the Schr\"odinger equation given in Eq. (\ref{se_E}) is valid for  {\em both} normal and binormal ribbons.
 Therefore, we will proceed to analyze this equation in terms of $E(\xi)$ using this general form valid for both ribbons.  In the final results we will substitute  the corresponding expressions  for $E(\xi)$ and $(M^2 -K)$ for the normal  ribbon from Eqs. (\ref{EFGn}) and (\ref{KnMn}), and for the binormal ribbon from Eqs. (\ref{EFGb}) and (\ref{KbMb}).
 
 Since the curved surface area element of the ribbon concerned is $da=h_1 ds d\xi=\sqrt{E(\xi)} ds d\xi$, a normalized wave function $\Phi(s,\xi)$  should satisfy
\be
\chi (s,\xi)= \dfrac{\Phi(s,\xi)}{\sqrt {h_1}}= \dfrac{\Phi(s,\xi)}{[E(\xi)]^{1/4}} ,
\label{chi-Phi}
\ee
so that $\int\int |\Phi|^{2} ds\,\, d\xi =1$.
Substituting Eq. (\ref{chi-Phi}) into Eq. (\ref{se_E}), a lengthy but straightforward calculation yields

 \be
 -\dfrac{\hbar^2}{2m} \dfrac{1}{E(\xi)} \dfrac{\partial ^{2} \Phi}{\partial s^{2}} -\dfrac{\hbar^2}{2m}  \dfrac{\partial ^{2} \Phi}{\partial {\xi}^{2}} + V_{eff} (\xi) \,\Phi = \mathcal{E} \,\Phi,
 \label{se_Veff}
 \ee
where  $V_{eff} (\xi)$ is given by 
\be
V_{eff} (\xi) =  - \dfrac{\hbar^2}{2m} \Big[\dfrac{3 (E_{\xi})^{2}}{16 E^{2}(\xi))}
   -\dfrac{E_{{\xi}{\xi}}} {4 {E(\xi)}}  + (M^2- K)\Big ].
\label{Veff}
\ee
Here $E_{\xi}$ and $E_{{\xi}{\xi}}$ represent the  first and second derivatives of $E$ with respect to $\xi$, respectively. After multiplying both sides of  Eq. (\ref{se_Veff})  by $E(\xi)$, we look for a separable solution of the form
\be
\Phi(s, \xi) = u(s) w(\xi) \,. 
\label{uw}
\ee
Dividing the resulting equation by  $u(s) w(\xi)$,
we get a  separable form for the Schr\"odinger equation, which enables us to write
\be
-\dfrac{\hbar^2}{2m}  \dfrac{\partial ^{2} u}{\partial s^{2}} = \mathcal{E}_{0}\, u(s) \,. 
\label{u}
\ee
Substituting Eq. (\ref{u})  in  the  separable form we obtained, multiplying the resulting equation by 
$w(\xi)/E(\xi)$ and rearranging the terms leads to the following Schr\"odinger equation for the wave function 
$w(\xi)$ 
\be
-\, \dfrac{\hbar^2}{2m}  \dfrac{\partial ^{2} w(\xi)}{\partial {\xi}^{2}} + \left[V_{eff}(\xi)  + \dfrac{\mathcal{E}_{0}} {E(\xi)}\right] w(\xi) = \mathcal{E} w(\xi).
\label{se_w}
\ee
Thus in the above Schr\"odinger equation,  the effective potential  is given by
\be
U_{eff} (\xi, {\mathcal{E}_{0}} )= \left[V_{eff}(\xi)  + \frac{\mathcal{E}_{0}} {E(\xi)}\right] \,, 
\label{tildeV}
\ee
where  $V_{eff}(\xi)$  is given in Eq. (\ref{Veff}).

\subsection{Effective quantum geometric potential: Helical  ribbon of finite Length}
\label{subsecEff}
 The effective potential  given by Eq. (\ref {tildeV}) depends upon the quantity $\mathcal{E}_{0}$
 which is to be found by solving Eq. (\ref{u}) for appropriate boundary conditions \cite{deLima}
on the wave function $u(s)$.  As should be clear, for  a particle confined  to a helical ribbon of length $L$,  the wave function $u(s)$  must vanish at the two ends of the ribbon, giving the boundary condition $u(0)= u(L) =0$. For the solution
 $u(s)= u_{0} \sin {\rm {k}}  s $, this condition yields 
 \be
 \rm {k}= \rm{k_{n}}  =  n\pi /L,
 \label{rmk}
 \ee
 where   $ n = 1, 2, 3...$. Hence only {\em nonzero} quantized values of $\rm {k}$  are allowed for both types of helical ribbons, implying  the existence of  {\em only moving particles}.
 
 A helical ribbon with length $L$  which is composed of integer $q$  complete  $2 \pi$  turns of the helix
 presents an interesting special case.  Here,
 $L=L (q) =  2 \pi q/ \sqrt{(k_{0}^2 +\tau_0^2)}$ (see Eq. (\ref{Lq})) so that  Eq. (\ref{rmk}) becomes
 \be
  \rm {k}={\rm k}_{n,\,q} = n\pi /L(q)= (n/2q) \sqrt{(k_{0}^2 +\tau_0^2)},
 \label{knq}
 \ee
 where   $n = 1, 2, 3...$ and $q=1,2,3...$.  For this case, the quantized value of $\rm {k}$ depends on  two  nonzero integers $n$ and $q$, leading to a rich behavior.
 We will therefore focus on the  effective quantum geometric potential for such ribbons.
 
  Thus  Eq. (\ref{u}) 
  yields $\mathcal{E}_{0} = \dfrac{\hbar^2 {\rm k} ^{2}}{2m} = \dfrac{\hbar^2 {\rm k_{n,q}} ^{2}}{2m}$.
Substituting  for     $\mathcal{E}_{0}$  in  Eq. (\ref {se_w}), the Schr\"odinger equation  for $w(\xi)$  becomes
\be
-\, \dfrac{\hbar^2}{2m}  \dfrac {\partial ^{2} w(\xi)}{\partial {\xi}^{2}} + U_{eff} (\xi, {\rm k_{n,q}}) w(\xi) = \mathcal{E} w(\xi) \,, 
\label{se_Vtilde}
\ee
where the  effective potential ${U_{eff}} (\xi, \rm{k_{n,q}})$ is given by
\be
U_{eff} (\xi, {\rm k_{n,q}})
 =  - \dfrac{\hbar^2}{2m} \Big[\dfrac{3 (E_{\xi})^{2}}{16 E^{2}(\xi))}
   -\dfrac{E_{{\xi}{\xi}}} {4 {E(\xi)}}  + (M^2- K) - \dfrac{{\rm k_{n,q}} ^{2}}{E(\xi)}  \Big ].
\label{Vqg}
\ee

 The  effective potential  given in (\ref {Vqg}) is  seen to be  purely quantum mechanical in origin  \cite{dandoloff,atanasov} since it is proportional to  
 $\hbar^{2}$.  In addition, it is also dependent on the {\em surface geometric parameters}  $E$  and 
 $(M^{2} - K)$  for  the respective ribbons. Hence  $U_{eff} (\xi, \rm{k_{n,q}})$ is called the {\em effective quantum geometric potential}. 

As  mentioned below Eq. (\ref{se_E}),  the Schr\"odinger equation, Eq.~(\ref{se_Vtilde}), is  
valid for {\em both}  the normal and binormal ribbons. To obtain their respective effective potentials,  the 
corresponding expressions for the geometric parameters for the normal and binormal ribbons should be substituted on the right hand side of   $U_{eff} (\xi, \rm{k_{n,q}})$ given in (\ref{Vqg}). 

For the normal and binormal ribbons, from Eqs. (\ref{EFGn})  and (\ref{EFGb}) we have
\be 
E^{(n)} (\xi) = [(1+k_0\xi)^2 + \tau_0^{2} \xi^{2}] = [1+ 2 k_0\xi + (k_0^2+\tau_0^2) \xi^2]\,\,\,;\,\,\,E^{(n)}_{\xi} = 2[k_0 + (k_0^2+\tau_0^2) \xi]\,\,\,;\,\,\, E^{(n)}_{\xi \xi} = 2 (k_0^2+\tau_0^2)
\label{dersEn}
\ee
and
\be
E^{(b)} (\xi) = (1+ \tau_0^{2} \xi^{2}) \,\,\,;\,\,\,E^{(b)}_{\xi} = 2\tau_0^2\, \xi\,\,\,;\,\,\, E^{(b)}_{\xi \xi} = 2 \,\tau_0^2 \,. 
\label{dersEb}
\ee
In Eq. (\ref{Vqg}), substituting  the expressions for $(M^{2}-K)$  given in the last entry of Eq. (\ref{KnMn}) [resp. Eq. (\ref{KbMb})] with
Eq. (\ref{dersEn}) [resp. Eq. (\ref{dersEb})]   for the normal [resp. binormal] helical ribbon,  
a  very lengthy calculation yields an effective quantum geometric potential  which  is given by the following  {\em same functional of $E(\xi)$} for  both the normal  and binormal helical ribbons.

\be
U_{eff}  (\xi, {\rm k_{n,q}}) = - \dfrac{\hbar^2}{8m} \Big[     
\dfrac{[   (k_0^{2}+\tau_{0}^{2}) -  4\rm{k_{n,q}}^{2}  ]}{E(\xi)}  +   \dfrac{\tau_0^{2}}{[ E({\xi})]^{2} }  \Big],
\label{VqgSame}
\ee
where ${\rm {k}_{n,q}}$ is defined in Eq. (\ref{knq}).

An inspection of our detailed calculations shows that this similarity in the  functional form  for {\em both} ribbons arises from the intricate interplay between  contributions to the quantum potential arising from the  (intrinsic) 
surface metric  part  $E$,  and the (extrinsic)  part [$(M^2-K)$]  [see  Eqs. (\ref{KnMn}) and (\ref{KbMb})] in the expression for the effective potential 
given in Eq. (\ref {Vqg}).


 Summarizing,  to find the effective  quantum geometric potentials  for the normal and binormal helical ribbons, we only need to substitute in Eq. (\ref{VqgSame}), the  respective expressions for $E(\xi)$ for these ribbons which are 
given in the first entries in Eqs. (\ref{dersEn}) and (\ref{dersEb}).  We note that 
 $E^{(n}({\xi})$ and  $E^{(b}({\xi})$ are positive for all $\xi$.   \\
 As is clear from Eq. (\ref {se_Vtilde}),   localized states are supported when the effective potential $U_{eff}  (\xi, \rm {k_{n,q}})$  given in  Eq. (\ref{VqgSame}) takes on {\em  negative}  values for all $\xi$. Additionally,
 Eq. (\ref{knq}) shows that $\rm {k_{n,q}}$ is always nonzero. This leads to the following  {\em geometric condition} on $\rm {k_{n,q}}$ for the existence of localized states:
 \be
 0 < \rm {k_{n,q}}^{2} \le \frac{(k_0^{2}+\tau_{0})^{2}}{4}.
\label{ineq}
\ee

We find it convenient to write Eq. (\ref{VqgSame}) in the form
\be
U_{eff}  (\xi, \beta_{n,q}) = - \dfrac{\hbar^2}{2m} \left[\dfrac{\beta_{n,q}\,(k_0^{2}+\tau_{0}^{2})}{E(\xi)} + \dfrac{\tau_0^{2}}{ 4 E^{2}({\xi})} \right] ,
\label{VqgSameBeta}
\ee
where the parameter $\beta_{n,q}$ is defined as
\be
\beta_{n,q} = \frac{1}{4} - \frac{{\rm {k_{n,q}}}^{2}}{(k_0^{2}+\tau_{0}^{2})}.
\label{beta}
\ee
Thus the condition for localized states given in Eq. (\ref{ineq})  becomes
$ 0\,\le\,\beta_{n,q} \,< \frac{1}{4}$.
 
 
Substituting Eq. (\ref{knq})  in Eq. (\ref{beta}) yields
\be
\beta_{n,q} = \frac{1}{4} \left( 1 - \frac{n^{2}}{q^{2}}\right).
\label{betanq}
\ee
 Using Eq. (\ref{betanq}), the quantum effective geometric potential given in Eq. (\ref {VqgSameBeta}) becomes
\be
U_{eff}  (\xi, n,q) = - \dfrac{\hbar^2}{8m} \left[\dfrac{\,(k_0^{2}+\tau_{0}^{2})\left( 1 - \frac{n^{2}}{q^{2}}\right)}{E(\xi)} + \dfrac{\tau_0^{2}}{  E^{2}({\xi})} \right] ,
\label{VqgSameBetanq}
\ee
where  $n=1,2, 3,....,q$ and  $q=1,2,3,...$.
Hence  the condition for  {\em localized states}  becomes
\be
1 \le  n^{2} \le  q^{2} \,. 
\label{ineqOpen}
\ee
 This gives the intriguing result that the particle momentum quantum number  for localized states is restricted 
 by the integer $q$. 
 
We note that  since the metric for the normal ribbon  $E^{(n)}({\xi})$ is greater than the metric for the binormal ribbon 
 $E^{(b)}({\xi})$,  
Eq. (\ref{VqgSameBetanq}) shows that these states satisfy
\be
U_{eff}^{({n})}  (\xi, n,q)\,\,  <  \,\,U_{eff}^{({b})}  (\xi, n,q).
\label{VnVb}
\ee

\subsection {Analysis of the effective  quantum geometric potential} 
\label{subsecAnaly} 

 We now analyze the expression for the effective quantum geometric potential $U_{eff}  (\xi, \beta_{n,q})$ 
 given in Eq. (\ref {VqgSameBeta}).  It is convenient  to use  the   notation $\beta$ for  
 $\beta_{n,q}$    in the following analysis. 
We define  
\be
C_0= 4\beta (k_0^{2}+\tau_{0}^{2}) \,. 
\label{Czero}
\ee
 We are mainly concerned with  {\em localized states} in this paper. For these states, $C_0\ge 0$, 
  as seen from Eq. (\ref {betanq}).
 Next, (setting the constant ${\hbar^{2}}/{2m} =1$ for convenience) we write the potential in Eq. (\ref{VqgSameBeta})  as
$U_{eff} (\xi)= -\dfrac{1}{4 E^{2}(\xi)}[C_{0}E(\xi) + \tau_{0}^{2}]$. This  yields  
\be
\dfrac {dU_{eff}}{d\xi} = \dfrac{1}{4 E^{3}(\xi)}E_{\xi}\,\,\Big[C_{0} E(\xi) + 2 \tau_{0}^{2}\Big],
\label {DerVeff}
\ee
where 
$E_{\xi} = {dE}/{d\xi}$.
Hence the extrema of $U_{eff}  (\xi, {\rm k})$  are given by either  (a) $E_{\xi} =0$ or
 (b)$ \Big[C_{0} E(\xi) + 2 \tau_{0}^{2}\Big] =0$.
Since $E(\xi)$ is a quadratic function of $\xi$ for both ribbons, Case (a)  leads to one extremum  and  Case (b) to two extrema. However, for localized states,  we  must have $C_0\ge 0$ as mentioned above. Further, 
 $E(\xi)$ is positive definite for both ribbons. Hence  the extremum condition for Case (b) cannot be satisfied for localized states.
  
Now, Case (a) corresponds to  $E_{\xi}=0$.
On using  Eqs. (\ref{dersEn})  and (\ref{dersEb}),  we find that  the  single extremum  for the normal  and binormal helical ribbon appears, respectively,  at
\be
\xi^{(n)} = \xi_{0}^{(n)}=  -\frac{k_{0}}{(k_0^{2}+\tau_{0}^{2})}\,\,\,\,; \,\,\,\,\xi^{(b)} = \xi_{0}^{(b)}=  0.
\label{NBmin}
\ee
Next,  we calculate
\be
\frac {d^{2}U_{eff}}{d\xi^{2}} = \frac{1}{4 E^{6}(\xi)}\,\Big[E^{3}(\xi)[C_{0} E(\xi) + 2 \tau_{0}^{2}] 
E_{\xi \xi} \,\,- \,\,2 E^{2} [C_{0} E(\xi) + 3 \tau_{0}^{2}] \,E_{\xi}^{2}\,\,\Big] \,. 
\label{Der2V}
\ee
On setting $E_{\xi}=0$ in Eq. (\ref{Der2V}), we find that  $\dfrac {d^{2}U_{eff}}{d\xi^{2}} > 0$, since $C_0 \ge 0$ for localized states. Hence the extremum    given in Eq. (\ref{NBmin}) is a {\em minimum} for both ribbons.

Although Case (b) is not relevant for the localized states we are concerned with, for completeness we also discuss this case. Here, the extremum condition is  $\Big[C_{0} E(\xi) + 2 \tau_{0}^{2}\Big] =0$, whose two roots  correspond to  two extrema. Let us denote them by $\xi_{\pm}$. Now, substituting this condition in Eq. (\ref{Der2V}),  we find  that at $\xi =\xi_{\pm}$, $\dfrac {d^{2}U_{eff}}{d\xi^{2}}= - \Big[ 2 E^{2}(\xi)  \tau_{0}^{2} \,E_{\xi}^{2}\,\,\Big] < 0$, showing that these two extrema will always be {\em maxima}. Our further analysis shows that two well-separated,  shallow  positive maxima of equal height appear at $\xi_{+}$ and ${\xi}_{-}$ for $\beta > 0$, which  {\em approach each other}, increasing  in height with increasing $n$ to merge into a single positive maximum.

The above analytical results agree with  Figs. 1 and 2, where we have
 plotted  the quantum effective potential $U_{eff}  (\xi, \beta_{n,q})$  given in
Eq. (\ref{VqgSameBeta}), as a function of $\xi$, for the normal and binormal ribbon, respectively.
 In Figs. 3 and 4,  we depict the normal and binormal helical ribbons, respectively. We find that the particle localizes near the inner edge for the normal helical ribbon, and on the central helix for the binormal helical ribbon, in agreement  with Figs. 1 and 2.
 
 \begin{figure}[t]
	\begin{center}
		\includegraphics[scale=0.4]{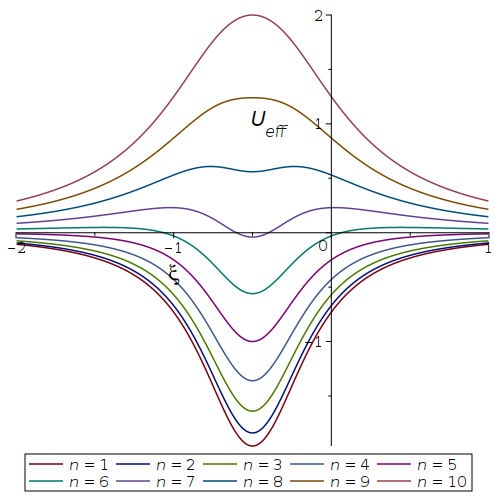}
		\caption{\label{fig:1}  Plots for the  effective  quantum geometric potential $U_{eff}  (\xi, \beta_{n,q})$  for the normal helical ribbon given in Eq. (\ref{VqgSameBeta}) [setting $\hbar^{2}/2m =1$] as a function of $\xi$, with $E(\xi) = E^{(n)}(\xi)$ given in Eq. (\ref{EFGn}). $\beta_{n,q} = \dfrac{1}{4} ( 1 - \frac{n^{2}}{q^{2}})$. We choose  $k_0=\tau_0=1$ and  $q=5$.  Plots from bottom to top  correspond to  moving particles  with quantum numbers 
$ n=  1,  2,....... 10$.
For  $n \le  5 $,  the moving  particle  localizes at the minimum given by 
$\xi_{min} = -k_0/(k_0^{2}+\tau_{0}^{2}) = -0.5$. For $n > 5$, two maxima start to appear on either side of the minimum, and approach each other as $n$ increases further to become a single maximum, as shown analytically in Sec. (\ref{subsecAnaly}).
		}
	\end{center}
\end{figure}

\begin{figure}[t]
	\begin{center}
		\includegraphics[scale=0.4]{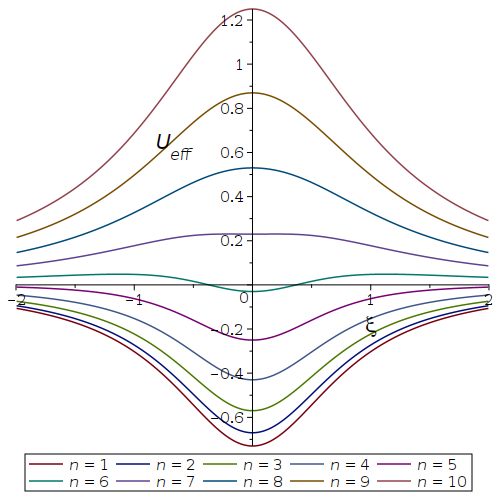}
		\caption{\label{fig:2} Plots for the  effective  quantum geometric potential $U_{eff}  (\xi, \beta_{n,q})$  for the binormal helical ribbon given in Eq. (\ref{VqgSameBeta}) [setting $\hbar^{2}/2m =1$] as a function of $\xi$, with $E(\xi) = E^{(b)}(\xi)$ given in Eq. (\ref{EFGb}). $\beta_{n,q} = \dfrac{1}{4} ( 1 - \frac{n^{2}}{q^{2}})$. We choose  $k_0=\tau_0=1$ and  $q=5$.  Plots from bottom to top  correspond to  moving particles  with quantum numbers $ n=  1,  2,..., 10$.
For $n \le  5 $,  the moving  particle  localizes at the minimum given by 
$\xi_{min} = 0$. For $n > 5$, two maxima start to appear on either side of the minimum, and approach each other as $n$ increases further, to become a single maximum, as shown analytically in 
Sec. (\ref{subsecAnaly}).
		 }
	\end{center}
\end{figure}

\begin{figure}[t]
	\begin{center}
		\includegraphics[scale=0.4]{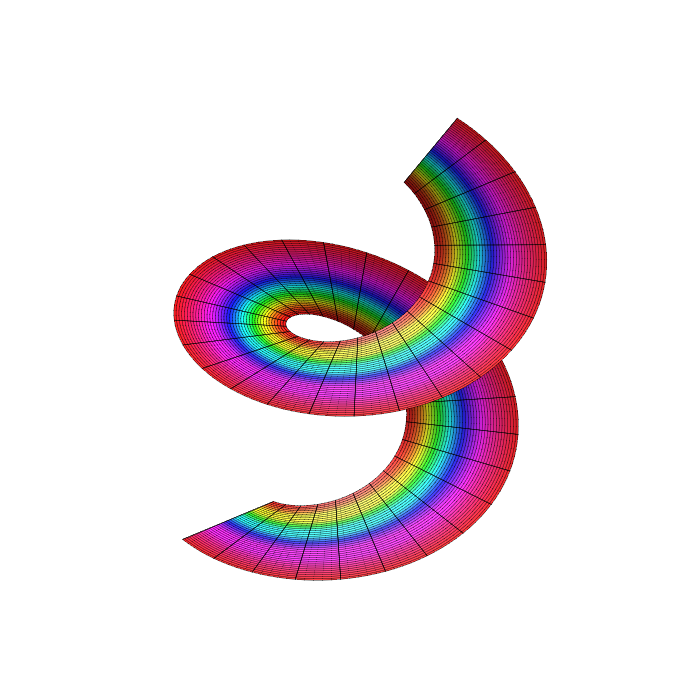}
		\caption{\label{fig:3} Normal helical ribbon. Width parameter $\xi$ for the ribbon is in the range $-0.5 < \xi < 0.33$. The coloring of the surface corresponds to the magnitude of the effective geometric potential  (see Fig.~1) for a quantum particle with  $n =1$. Only one turn of the ribbon is displayed. Particle localization, which corresponds to the minimum of the potential, is near the inner edge of the ribbon as highlighted in red. This is in agreement with Fig.~1.
		}
	\end{center}
\end{figure}

\begin{figure}[t]
	\begin{center}
		\includegraphics[scale=0.4]{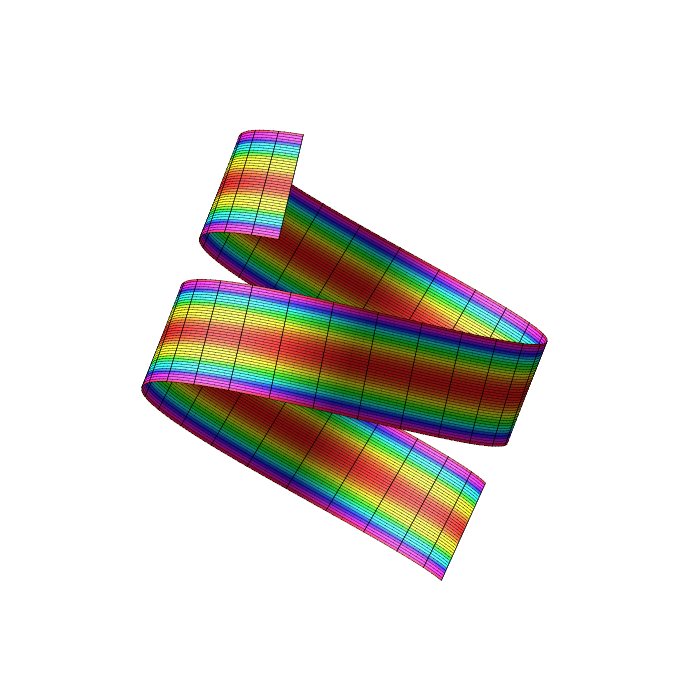}
		\caption{\label{fig:4} Binormal helical ribbon. Width parameter $\xi$ for the ribbon is in the range 
		$-0.33 < \xi < 0.33$. The coloring of the surface corresponds to the magnitude of the effective geometric potential  (see Fig.~2 ) for  a quantum particle with $n=1$. Only one turn of the ribbon is displayed. Particle localization, which corresponds  to the minimum of the potential,  is on the central helix of the ribbon as highlighted in red. This is in agreement with Fig.~2.
		}
	\end{center}
\end{figure}

\section {Quantum analog of  the Coriolis effect  in a helical nanoribbon}
\label{Qanalog}
 As analytically shown in Sec. (\ref{subsecAnaly}) and illustrated  in Figs. 1 and 2, a single minimum of the effective quantum geometric potential  occurs at
$\xi =\xi^{(n)}_{min}=  -\frac{k_{0}}{(k_0^{2}+\tau_{0}^{2})}$  for the normal helical ribbon, and at $\xi =\xi^{(b)}_{min}= 0$, for the binormal helical ribbon.  Further,  it was shown in In Sec. (\ref{subsecEff})  that for a quantum particle confined to a helical ribbon of any finite length, the wave vector ${\rm k}$ takes on quantized  {\em nonzero} values. Thus we have only a moving particle on such a ribbon.
  
  Such a localization of a moving particle on a helical curve at a specific width  $\xi=\xi_{min}$ for each helical  ribbon suggests the presence of  a pseudo-force acting on the particle (with wave function $u(s)$ satisfying Eq. (\ref{u}))  moving  along the tangent vector ${\bf X}_{s}(s,\xi)$ to the $s$-curve  on the surface of the ribbon, which  deflects  the particle transversely along the width $\xi$ of the ribbon concerned. 
 (Transverse, because  ${\bf X}_s.{\bf X}_{\xi} =0$ for  both types of  helical ribbons.) 
 
 This  force is highly reminiscent of
the  fictitious  Coriolis force  ${\bf F_{cor}}$  on a classical particle  moving with velocity ${\bf v}$  in a  non-inertial frame rotating with an angular velocity ${\boldsymbol \Omega}$ with respect to an  inertial frame, which transversely deflects the particle. For a particle with mass $m$, it  is given by \cite{goldstein}
\be
{\bf F}_{cor} =  -2m \,{\boldsymbol \Omega} \times {\bf v} \,. 
\label{cor}
\ee

In what follows, we investigate whether the   quantum particle localization  at specific widths for each ribbon, which is essentially  caused  by the  quantum geometric potential $U_{eff}  (\xi, \beta)$  given in Eq. (\ref{VqgSameBeta}) 
can be cast in the  form  given in Eq. (\ref{cor}), so that this phenomenon  can be regarded  as a quantum analog of the (classical) Coriolis effect \cite{gardas}. 

\subsection {General methodology applicable to both types of ribbons}

Our aim is to show that an equation analogous to Eq. (\ref{cor}) can be written down for a quantum particle moving  on a helical ribbon. We accomplish this by identifying  the quantum analog of each of the terms  appearing in that equation. 

We have seen that  a helical ribbon of any finite length supports {\em only} moving quantum particles. We find the quantum analog of  $m {\bf v}$ on a helical ribbon  appearing on the RHS of  Eq. (\ref{cor}) as follows.

 Let  ${\bf X}$   denote the surface of a helical ribbon. (See Eqs. (\ref{n}) and (\ref{b}).) 
 Let the  quantum particle satisfying Eq. (\ref{u}) move along the  helical $s$-curve at a given point $\xi$  lying on the width of the ribbon,  with velocity ${\bf v}$.  As should be clear,
 \be
{\bf v} = v \,\,\frac{{\bf X}_s}{|{\bf X}_s|} = \, v\,\hat{{\bf T}}(s,\xi),
\label{v}
\ee
where 
\be
\hat{{\bf T}}(s,\xi)  = \frac{{\bf X}_s}{|{\bf X}_s|}
\label{T}
\ee
is the unit tangent vector along the $s$-curve at a specific width coordinate $\xi$.

Using the well known de Broglie relation $mv = \hbar \, \rm {k}$, we  obtain the {\em quantum analog}  of $m {\bf v}$ appearing on the RHS of Eq. (\ref{cor}) as
\be
 m {\bf v} = \hbar\,\rm{k}\,\, \hat{{\bf T}}(s,\xi).
 \label{mvQ}
 \ee

 Since we  are only concerned with {\em localized states}  for the wave function, we are justified in invoking the Ehrenfest theorem \cite{messiah}  to write  the  {\em quantum analog}  of the force deflecting the quantum particle transversely on  a helical ribbon  as the negative gradient of the   corresponding potential function. Thus we obtain
 \be
{\bf F}_{Q} \,=\,- [d U_{eff}  (\xi, \beta)/d\xi] \,\,\hat{\boldsymbol \xi} ,
\label{FQ}
\ee
 where  the effective quantum geometric potential  $U_{eff}  (\xi, \beta)$ is given in 
 Eq. (\ref{VqgSameBeta}). 
  In  Eq. (\ref{FQ}), we have also used the fact that  the direction of the force  that pushes the particle  along the  width of the ribbon is just the  `unit width vector' $ \hat{{\boldsymbol \xi}}= {\boldsymbol \xi}/|{\boldsymbol \xi}| $ in the  ${\boldsymbol \xi}$ direction. From the definition of the two ribbons given in Eqs. (\ref{n}) and (\ref{b}), 
we see that ${\boldsymbol \xi}$ is along ${\bf n}$ and ${\bf b}$ respectively, for the normal and binormal ribbons. 

From  Eq. (\ref{DerVeff}), we write  the following general expression for $[d U_{eff}  (\xi, \beta)/d\xi]$ appearing in Eq. (\ref{FQ})  as 
\be
 - [d U_{eff}  (\xi, \beta)/d\xi] = -[\hbar^{2}/2m] G(\xi, \beta))  (dE/d\xi),
 \label{DerVeffQ}
 \ee
 where  $G(\xi, \beta)$ is given by
 \be
G (\xi, \beta)= [C_0 E(\xi) + 2 \tau_{0}^{2}] / 4 E^{3}(\xi) = [ 4\beta (k_0^{2}+\tau_{0}^{2}) \,E(\xi) +2 \tau_{0}^{2}] / 4 E^{3}(\xi).
 \label{G}
\ee
In the above, we have substituted for $C_0$ from Eq. (\ref{Czero}).  From Eq. (\ref{beta}), we have  $\beta = \frac{1}{4} - \frac{{\rm {k}}^{2}}{(k_0^{2}+\tau_{0}^{2})}$, on suppressing  the quantum indices $n$ and $q$ on $\beta$ and $\rm{k}$ for convenience. Substituting for $\beta$ we get
\be
G (\xi, \beta)= G(\rm{k},\xi)= \left[(k_0^{2}+\tau_{0}^{2})\left(1 -\, \frac{4{\rm{k}}^{2}}{(k_0^{2}+\tau_{0}^{2})}\right)\,\,E(\xi) +2 \tau_{0}^{2}\right] / 4 E^{3}(\xi).
\label{Gk}
\ee
In Eq. (\ref{DerVeffQ}),
we have re-introduced the factor $[\hbar^{2}/2m]$ which had  earlier been set to unity in the potential  in Sec. (\ref{subsecAnaly}) for calculational convenience. As is clear,
Eq. (\ref{DerVeffQ}) is valid for both ribbons, with the appropriate expressions for $E(\xi)$ substituted
for the respective ribbons.

Substituting Eq. (\ref{DerVeffQ}) in Eq. (\ref{FQ}) 
we obtain the  quantum analog of the force that is pushing the particle in the transverse direction  to be 
\be
{\bf F}_{Q} \,=\,-[\hbar^{2}/2m] G(\rm{k},\xi)  (dE/d\xi)\,\,\hat{\boldsymbol \xi} \, 
\label{FQG}
\ee
where $G(\rm{k},\xi)$ is given in Eq. (\ref{Gk}).

In order to interpret Eq. (\ref{FQG})   as a quantum analog  of  the Coriolis force,  we should be able to write it in the form  of Eq. (\ref{cor}) and find the quantum analog  of the angular velocity ${\boldsymbol \Omega}$.  To this end, we  first substitute
 the quantum analog  of   $m {\bf v}$  given in Eq. (\ref {mvQ}), along with the quantum analog  ${\bf F_{Q}}$  given in  Eq. (\ref{FQG})  into Eq. (\ref{cor}).  We then proceed to  identify the {\em quantum analog}  of the angular velocity  denoted by ${\boldsymbol \Omega}_{Q}$ in such a way that it satisfies
 \be
{\bf F_{Q}} \,=\,-[\hbar^{2}/2m] G(\rm{k},\xi)  (dE/d\xi)\,\,\hat{\boldsymbol \xi} = {\boldsymbol \Omega}_{Q}\, \times  -2\hbar\,\rm{k}\,\, \hat{{\bf T}}(s,\xi).
\label{FQanalog}
\ee
This yields
\be
 [\hbar/(4m {\rm k})] G(\rm{k},\xi)  (dE/d\xi)\,\, 
\hat{\boldsymbol \xi} \, =  \, {\boldsymbol \Omega}_{Q} \,\,\times \,\,\hat{{\bf T}}(s,\xi) 
\label{omegaQ}
\ee
where $G(\rm{k},\xi)$ is given in Eq. (\ref{Gk}).
In the above,  the quantities $\hat{{\bf T}}(s,\xi)$, $G(\rm{k},\xi)$, $E(\xi)$  and  $\hat{{\boldsymbol \xi}}$ are  ribbon-dependent, and will be explicitly computed for the normal and binormal helical ribbons.
 Using them in Eq. (\ref{omegaQ}), enables us  to   identify   the {\em quantum analogs}  (${\boldsymbol \Omega}_{Q}$) of the respective  angular velocities  of the   rotating (non-inertial ) Coriolis frames for the  two types of ribbons, as we show below. 
 
 \subsection  {Angular velocities of rotating frames for normal and binormal helical ribbons}
  We will use the superscripts $(n)$ and $(b)$ to denote quantities pertaining to the normal and binormal helical ribbons respectively.
 
 \noindent {\bf {\em Normal helical ribbon}:} From  the general expression  for $\hat{{\bf T}}(s,\xi)$ given in (\ref{T}),  a short calculation using  the definition Eq. (\ref{n}), along with the Frenet-Serret equations, Eq. (\ref{FrS}), for the helix yields
\be
\hat{{\bf T}}^{(n)}(s,\xi) = \frac{{\bf X}_{s}^{(n)}}{|{\bf X}_{s}^{(n)}|}  = 
 \big[(1 + k_{0} \xi) {\bf t} - \tau_{0}\xi {\bf b}\big ]/ E^{(n)} (\xi) \,. 
\label{Tn}
\ee
 Since $\hat{{\boldsymbol \xi}} ={\bf n}$ for the normal ribbon,  the LHS of Eq. (\ref{omegaQ}) is  a vector along ${\bf n}$.  The vector $\hat{{\bf T}}^{(n)}(s,\xi)$ appearing on its RHS is given in Eq. (\ref{Tn}). This
 suggests the natural choice
\be
{\bf \Omega}_{Q}^{(n)} = \Omega_{Q}^{(n)}  \hat{\bf e}_{z},
\label{ez}
\ee
 where  $\hat{\bf e}_{z} =[0,0,1]$ is a unit vector along the $z$-axis. 
Indeed,  on using Eqs. (\ref{Vectktau}) and (\ref{Vecnbktau}), a simple calculation yields
\be
\hat{\bf e}_{z}\, \times \,{\bf t} = (k_{0}/\alpha)\,{\bf n}\,\,\,\,;\,\,\,\,\hat{\bf e}_{z}\, \times \,{\bf b} =-(\tau_{o}/\alpha)\, {\bf n},
\label{ezcross}
\ee
where $\alpha= \sqrt {(k_{0}^2 + \tau_{0}^2)}$.  
On using  Eqs. (\ref{ez}) and  (\ref{Tn}), along with Eq. (\ref{ezcross}), a short calculation shows that  the RHS of Eq. (\ref{omegaQ}) becomes
\be
{\bf \Omega}_{Q}^{(n)} \,\,\times \,\,\hat{{\bf T}}^{(n)}(s,\xi)\,=\, \frac{\Omega_{Q}^{(n)} }{ E^{(n)}(\xi)\sqrt{(k_{0}^2 + \tau_{0}^2)}}\,\,[k_{0} + (k_{0}^2 + \tau_{0}^2)\xi] \,\, {\bf n}.
\label{RHSOmegaQn}
\ee
Since  $\hat{\boldsymbol \xi}$ is along the normal ${\bf n}$,   substituting Eq. (\ref{RHSOmegaQn}) in
 Eq. (\ref{omegaQ}), and  using   $(dE^{(n)}/d\xi) = 2 [k_{0} + (k_{0}^2 + \tau_{0}^2)\xi]$  [from Eq. (\ref{EFGn})]  on its LHS,  a short calculation yields
\be
\Omega_{Q}^{(n)}(\rm{k},\xi)\,=\,\frac{\hbar}{2m {\rm k}}\,\,\,\sqrt{(k_{0}^2 + \tau_{0}^2)}\,\, E^{(n)}(\xi)\,\,G^{(n)}(\rm{k},\xi).
 \label{omeganG}
 \ee
 Using the above expression, the angular velocity vector  for the normal ribbon is given by ${\bf \Omega}_{Q}^{(n)} = \Omega_{Q}^{(n)}(\rm{k},\xi)  \hat{\bf e}_{z}$, as seen from Eq. (\ref{ez}). Hence its axis is fixed along the $z$-axis.
 
 \noindent {\bf {\em Binormal helical ribbon:}}  Here, proceeding as in the case of the normal ribbon,  using  
 Eq. (\ref{b}) in Eq. (\ref{T}) yields
 \be
\hat{{\bf T}}^{(b)}(s,\xi) = \frac{{\bf X}_{s}^{(b)}}{|{\bf X}_{s}^{(b)}|}  = ({\bf t} - \tau_{0}\xi {\bf n})/ E^{(b)}(\xi) \,.
\label{Tb}
\ee
Since $\hat{{\boldsymbol \xi}} ={\bf b}$ for the binormal ribbon,  the LHS of Eq. (\ref{omegaQ}) is  a vector along ${\bf b}$, 
 suggesting  the natural choice
 \be
{\bf \Omega}_{Q}^{(b)} = -\Omega_{Q}^{(b)} \,\,{\bf n}.
\label{exy}
\ee
From  Eq. (\ref{Vecnbktau}), we find that
 $\bf {n}$  is  a unit vector in the $(x,y)$ plane. 
On using  Eqs. (\ref{exy}) and  (\ref{Tb}), the RHS of Eq. (\ref{omegaQ}) yields
\be
{\bf \Omega}_{Q}^{(b)} \,\,\times \,\,\hat{{\bf T}}^{(b)}(s,\xi) \,= \,  [\Omega_{Q}^{(b)}/ E^{(b)}(\xi)]\,\,{\bf b}.
\label{RHSOmegaQb}
\ee
 Since  $\hat{\boldsymbol \xi}$ is along the binormal ${\bf b}$,  substituting $(dE^{(b)}/d\xi) = 2\,\tau_{0}^2 \xi$
   [from Eq. (\ref{EFGb})]  on the LHS of Eq. (\ref{omegaQ}) and simplifying, we find
\be
\Omega_{Q}^{(b)}(\rm{k},\xi)\,=\,\frac{\hbar}{2m {\rm k}}\,\,\tau_{0}^2 \,\, E^{(b)}(\xi)\,\,G^{(b}(\rm{k},\xi)\,\,\xi.
 \label{omegabG}
 \ee
Using the above expression, the angular velocity vector  for the binormal ribbon is given by ${\bf \Omega}_{Q}^{(b)} = \Omega_{Q}^{(b)}(\rm{k},\xi) \,{\bf n}$, as seen from Eq. (\ref{exy}). This shows that  the  axis of the angular velocity ${\bf \Omega}_{Q}^{(b)}$ of the rotating frame for the binormal ribbon is itself rotating  in the $(x,y)$ plane, indicating precession. The physical interpretation for the appearance of the linear term $\xi$  in Eq. (\ref{omegabG}) will be given in the next section.

We consider ribbons of length $L(q)$ composed of integer  $q$  turns. As we have shown in  Eq. (\ref{knq}),
 ${\rm k}$ takes on  {\em quantized}  nonzero values ${\rm k}={\rm k}_{n,q} =  (n/2q) \sqrt{(k_{0}^2 +\tau_0^2)}$, $n= 1, 2, 3,...,q$, with $q =1,2,3...$. Note that in Eqs. (\ref{omeganG}) and (\ref{omegabG}),  the  expressions for $G^{(n}(\rm{k},\xi)$ and   $G^{(b}(\rm{k},\xi)$  are obtained  by setting $E(\xi)= E^{(n)}(\xi)$ and $E(\xi)= E^{(b)}(\xi)$ respectively, in Eq. (\ref{Gk}), and substituting the quantized values of  $\rm{k}$ in them. Simplifying, 
 Eq. (\ref{omeganG}) yields
\be
\Omega_{Q}^{(n)} (n,q,\xi)\,=\,\frac{\hbar}{m}\,\,\frac{q}{n} \,\, \frac{\left[ (k_{0}^2 +\tau_0^2) \left(1-\dfrac{n^{2}}{q^{2}}\right)\,E^{(n)}(\xi) + 2 \tau_{0}^{2}\right] }{ [E^{(n)}(\xi)]^{2}}
\label{omegannqGen}
\ee
 for the normal ribbon.
 Likewise, Eq. (\ref{omegabG}) gives
 \be
\Omega_{Q}^{(b)}(n,q,\xi) \,= \,\frac{\hbar}{m} \frac{q}{n} \,\,\frac{\tau_{0}^{2}}{\sqrt{(k_{0}^2 +\tau_0^2)}}\, \frac{[(k_{0}^2 +\tau_0^2) (1-\frac{n^2}{q^{2}}) E^{(b)}(\xi) + 2 \tau_{0}^{2}] }{[E^{(b)}(\xi)]^{2}}\,\,\,\,\xi
\label{omegabnqGen}
\ee
 for the binormal ribbon.
  
 Note that  $k_{0}$ and $\tau_{0}$ are nonzero constants, and $n$ is nonzero.
 Hence the quantum analogs of the angular velocities  we have found  in Eqs. (\ref{omegannqGen}) and 
 (\ref{omegabnqGen}) are well-defined quantities.

 Specializing to  {\em long ribbons } with  number of turns $q \gg1$  the angular velocity for the  low momenta localized states with $n \ll q$, the term $(1-\dfrac{n^{2}}{q^{2}})$ goes to unity. Hence for both the ribbons, the angular velocity becomes proportional to $\dfrac{q}{n}$ in this case. Therefore as $n$ increases, the quantum effect gets less pronounced, as expected physically.

\subsection  {Comparison of rotating frame angular velocities obtained for normal and binormal helical ribbons}

It is instructive to present the physical interpretation of the expressions for the quantum analog of the Coriolis rotating frame angular velocities  obtained for the  normal and binormal helical ribbons, respectively, and  compare them. 

Firstly, the appearance of $\hbar$ in the general expressions for the angular velocity  of the frame of rotation for the normal  helical ribbon 
$\Omega_{Q}^{(n)}(\rm{k},\xi)$ and binormal helical ribbon $\Omega_{Q}^{(b)}(\rm{k},\xi)$, given in  Eqs.  (\ref {omeganG}) and (\ref{omegabG}) respectively, shows that they are indeed quantum analogs.

 

On a helical  ribbon, for every width point $\xi$, there is a helix, and a quantum particle moving on the helix with a momentum $\rm{k}$ experiences a push that depends on $\xi$. Hence the quantum analog of the angular velocity of the rotating frame that causes the Coriolis effect  depends on $\rm{k}$ and $\xi$, as expected.

The dependence of the angular velocities  on  the width coordinate $\xi$ is {\em distinct} for the two ribbons, as seen from the general expressions given in
  Eqs.  (\ref {omeganG}) and (\ref{omegabG}). This can be understood as follows.
Note that  in  $\Omega_{Q}^{(b)}(\rm{k},\xi)$, the linear dependence on $\xi$ implies that  the force on the particle on a binormal ribbon acts in opposite directions for $\xi <0$ and $\xi>0$.  This leads  to quantum localization at the central helix on the binormal helical ribbon as required. This is unlike for the normal ribbon where  $\Omega_{Q}^{(n)}(\rm{k},\xi)$ does not change its sign  as $\xi$ changes sign, since $E^{(n)}(\xi)$ is positive for both signs of $\xi$, as seen from Eq. (\ref{EFGn}). This results in localization near the inner edge of the normal ribbon as needed. Thus, the expressions of the angular velocities obtained  are consistent with  our results on quantum localization.

 The  angular velocity of the rotating frame  in the Coriolis effect is taken to be a constant  in most books on classical mechanics \cite{goldstein}. However, the quantum expressions in   Eqs.  (\ref {omeganG}) and (\ref{omegabG}) are dependent on  the wave vector $\rm k$, which essentially translates to the  velocity of the particle in the classical scenario. Such a dependence is  not unphysical. For example,  angular velocities of rotation in some fluids depend on the particle flow velocity \cite{bush}.  Such dependencies can also appear if the underlying dynamics of the physical system considered is nonlinear \cite{strogatz}.

 
   Note that the derivation of the  angular velocity  of the  Coriolis rotating frame for a helical ribbon uses the Frenet-Serret equations, which  in turn represent the  rotating frame of the Frenet triad, as explained below Eq. (\ref{FrS}). This  shows an implicit connection between these two rotating frames.
  
 
 In summary, {\em given} the quantum geometric potential that is the cause of the localization of a moving particle on the width of the ribbon, suggesting the presence of a pseudo-force, our aim was to {\em find} a consistent quantum analog of the angular velocity ${\bf \Omega}$ for the rotating frame on the ribbon in such a way that this phenomenon can be interpreted as the quantum analog 
 of the Coriolis effect. We have achieved that.

\section{Discussion}
Most of the results obtained in this paper have been summarized in the Introduction. 
It is important to note that  geometric effects on quantum particle transport (induced by the curvature and torsion of  helical ribbons) which  we have studied in this paper are much deeper and much more  general than this particular example, with  its roots in  the concept of parallel transport of a vector around a closed path on a curved manifold and  the associated anholonomy. This leads  \cite{berry1} to the appearance of  a  Berry connection as well as  a Berry curvature, representing the  analogous vector potential and its  corresponding divergence-free  `magnetic field', respectively.  In addition,  the integral of the Berry connection leads to a Berry phase for the wave function. These have appeared in diverse contexts in physics \cite{berry2}.

Parallel transport of a vector on a  curved  surface  was analyzed  over three decades ago \cite{balakrishnan1}, by regarding the surface as an evolving space curve. By using the Darboux rotation of the  Frenet-frame,   a general expression for the  anholonomy  that arises for a closed path on the curved surface was derived,  with the emergence  of a  vector potential and a geometric phase.  Non-commutative geometry is inherent in the quantum phenomena  that arise in systems associated with a vector potential. Recently, topological invariants and their underlying geometric structure have been considered in general \cite{balakrishnan2}. We plan to apply these results to helical ribbons in a separate paper.

  Our continuum analysis is valid in the ballistic regime,  which is accessible in experiments  on nanoribbons \cite{baringhaus}. The methodology we have presented  for helical ribbons can also be extended to other curved surfaces. Diverse devices that are applications of the effect of geometry
 on quantum transport,  controlled by appropriately tuning  the curved surface geometry of the surface considered, can be envisaged in nanotechnology and biotechnology.  As our results show, the interplay between quantum mechanics and curved geometry  can be experienced  not only by electrons, where it manifests itself as a Hall-like effect on helical ribbons, but  by  other quantum particles as well, under suitable scenarios.
 
 Quantum mechanics  of particles on  curved manifolds can lead to new phenomena. As mentioned in the Introduction, it has recently  been noted \cite{agarwal}  that   in the case of a three dimensional spirally stacked tungsten disulfide  (WS$_2$),  “the twisting structure and the Coriolis-like force were guiding the electrons”, leading to a `twistronic Hall effect' \cite{ji}.
  This system is different from the helical ribbons we have studied in the present paper.
  We hope that our work will motivate  experiments to study  transport  of quantum particles on  normal and binormal helical nanoribbons to look for  novel effects.

 \section{Acknowledgments} 
The work  of A.S.  at Los Alamos National Laboratory was carried out under the auspices of the U.S. DOE and NNSA under Contract No. 89233218CNA000001.

\end{document}